\documentstyle[epsfig]{aipproc}

\begin{document}
\title{ Why $T_c$ is too high when antiferromagnetism is underestimated?\\ 
--- An understanding based on the phase-string effect}

\author{Z. Y. Weng}
\address{Texas Center for Superconductivity, University of Houston\\
Houston, TX 77204-5932\\
E-mail: zyweng@uh.edu}

\maketitle

\begin{abstract}
It is natural for a Mott antiferromagnetism in RVB description 
to become a superconductor in doped metallic regime. But the issue of
superconducting transition temperature is highly nontrivial, as the AF fluctuations in the form of RVB pair-breaking are crucial in 
determining the phase coherence of the superconductivity. Underestimated AF 
fluctuations in a fermionic RVB state are the essential reason causing an 
overestimate of $T_c$ in the same system. We point out that by starting with a 
{\it  bosonic} RVB description where both the long-range and short-range AF 
correlations can be accurately described, the AF fluctuations can effectively reduce $T_c$ to a reasonable value through the phase string effect, by 
controlling the phase coherence of the superconducting order parameter.  
\end{abstract}

It was first conjectured by Anderson\cite{anderson} that the ground state of the two-dimensional (2D) $t-J$ model 
may be described by some kind of resonating-valence-bond (RVB) state. Perhaps the most natural 
consequence of a RVB description is the superconductivity once holes are 
introduced into the system, which otherwise is a Mott insulator, as preformed 
spin pairs become mobile, i.e., carrying charge like Cooper pairs.

Even though the RVB state was proposed\cite{anderson} to explain the 
then-newly-discovered high-$T_c$ superconductor in cuprates, the 
mean-field estimate of $T_c$ turned out to be way too high ($\sim 1000$$ K$ at 
doping concentration $\delta\sim 0.1$\cite{nagaosa}) as compared to the 
experimental ones ($\sim 100$$K$). Another drawback for the earlier fermionic RVB
theory (where spins are in fermionic representation) is 
that the antiferromagnetic (AF) correlations are always underestimated,
which becomes obvious in low-doping limit where long-range AF ordering (LRAFO) cannot be naturally recovered. Even at finite-temperature where the LRAFO is absent, the spin-lattice relaxation rate calculated based on those RVB theories shows a wrong temperature-dependence as compared
to that well-known for the Heisenberg model, indicating an absence of 
AF fluctuations. 

Intuitively, a fermionic RVB state should become superconducting of 
BCS type at finite doping where the RVB pairs are able to move around carrying charge. But since the RVB pair-breaking process corresponds to AF
fluctuations at insulating phase while it also represents Cooper pair-breaking in superconducting state,  it is not difficult to see why the underestimate of AF fluctuations in the fermionic RVB theory would be generally related to an 
overestimated $T_c$. 

Of course, the above-mentioned drawback in describing antiferromagnetism does 
not include all RVB theories. There actually exists a RVB state which can describe the AF correlations
extremely well. As shown by Liang, Douct, and Anderson\cite{liang}, the trial wavefunction of RVB spins in {\it bosonic} representation can reproduce almost exact ground-state energy at half-filling (which implies 
a very accurate description of short range spin-spin correlations). A simple mean-field theory of bosonic 
RVB studied by Arovas and Auerbach\cite{aa} (usually known as the Schwinger-boson
theory) can easily recover the LRAFO at zero-temperature and reasonable behavior of magnetic
properties at finite temperature. 

Thus, one may classify two kinds of RVB states based on whether the 
spins are described in fermionic or bosonic representation. In 
principle, both representations should be equivalent mathematically
due to the constraint that at each site there can be only one spin. But 
once one tries to do a mean-field calculation by relaxing such a constraint, two representations will 
result in qualitatively different consequences: in fermionic representation, even an exchange of two 
spins with the same quantum number will lead to a sign change of the
wavefunction as required by the fermionic statistics. At half-filling, this is apparently redundant as the 
true ground-state wavefunction only changes sign when two opposite spins at different sublattice sites exchange with each other, known as the Marshall sign\cite{marshall} which can be easily incorporated into the bosonic RVB description. That explains the great success 
of the bosonic RVB mean-field theory over the fermionic ones at half-filling.

Since the bonsonic RVB description of antiferromagnetism is proven strikingly accurate at half-filling,
one may wonder why we cannot extend such a formalism to the doped case by literally doping
the Mott-insulating antiferromagnetism into a metal (superconductor). In fact, people have tried this
kind of approach based on the so-called slave-fermion representation but the mean-field theories always
lead to the so-called spiral phase\cite{ss} which is inherently unstable against the charge fluctuations\cite{weng0}. In other words, an
instability boundary seems to prevent a continuous evolution of the mean-field bosonic RVB description
into a short-ranged spin liquid state at finite doping.

It implies that some singular effect must have been introduced by doping which was overlooked in those
theories. Indeed, it was recently revealed\cite{weng1} that a hole hopping on the antiferromagnetic background always leaves a string 
of phase mismatch (disordered Marshall signs) on the path which is non-repairable at low-energy (because
the spin-flip term respects the Marshall sign rule). The implication of the existence of phase string
is straightforward:  a hole going through a closed loop will acquire a nontrivial Berry phase and a 
quasiparticle picture no longer holds here. This explains why the mean-field theory in 
the slave-fermion representation, where the topological effect of the phase string is smeared out by
mean-field approximation, always results in an unphysical spiral-phase instability.  

This barrier can be immediately removed once the nonlocal phase string effect is explicitly incorporated
into the Schwinger-boson, slave-fermion representation through a unitary transformation -- resulting in
the so-called phase string formulation\cite{weng2} where the mean-field treatment generalized from the 
Schwinger-boson mean-field state at half-filling\cite{aa} becomes workable at finite doping. A metallic phase\cite{weng3} with 
short-range spin correlations can be then obtained in which the ground state is, not surprisingly, always superconducting. 

What becomes special here is that the phase string effect introduces a 
phase-coherence factor to the superconducting order parameter\cite{weng3}:
\begin{equation}\label{1}
\Delta^{SC}_{ij}\propto \rho^0_s\Delta^s e^{\frac{i}{2}(\Phi^s_i+\Phi^s_j)}
\end{equation}
where $\Delta^s$ denotes the mean-field RVB order parameter for bosonic spinons and $\rho^0_s
\sim \delta$ is the bare superfluid density determined by holons (both spinon and holon are bosonic
in this representation), $i$ and $j$ refer to two nearest-neighbor sites. The phase-coherence factor
$e^{\frac i 2 \Phi^s_i}$ is related to the spin degrees of freedom as follows
\begin{equation}\label{2}
\Phi_i^s= \sum_{l\neq i} \mbox{Im ln $(z_i-z_l) \sum_{\alpha}\alpha n^b_{l\alpha}$}
\end{equation}
with $n^b_{l\alpha}$ being defined as the spinon number operator at site $l$. The physical interpretation 
of the phase-coherence factor (\ref{2}) is that each spinon contributes to a phase-vortex (anti-vortex). 

\begin{figure}[hbp!] 
\centerline{\epsfig{file=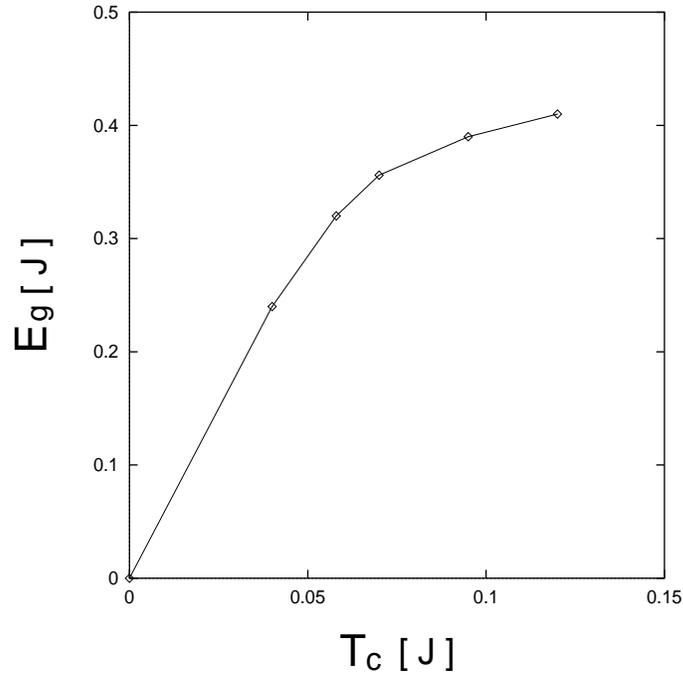,height=3.5in,width=3.5in}}
\vspace{10pt}
\caption{The relation of $T_c$ with the spin characteristic energy $E_g$
defined in Fig. 2}
\label{fig1}
\end{figure}
At zero 
temperature, when all spinons are paired, so are those vortices and anti-vortices, such that 
superconducting order parameter $\Delta^{SC}$ achieves 
phase-coherence\cite{emery}. At finite temperature,
free excited spinons or dissolved vortices (anti-vortices) tend to induce a 
Kosterlitz-Thouless type transition once
the ``rigidity'' of the condensed holons breaks down which may be estimated as the excited spinon
number becomes comparable to the holon number\cite{weng3}. 

The superconducting transition temperature obtained this way is shown in Fig. 1 versus a spinon 
characteristic energy scale $E_g$. The definition of $E_g$ is shown in Fig. 2 where the local
(${\bf q}$-integrated) dynamic spin susceptibility as a function of energy is given at $\delta=0.143$
(solid curve) at zero temperature. As compared to the undoped case ($\diamond$ curve), a resonance-like
peak emerges at low-energy $E_g$ due to the phase string effect. The doping-dependence of $
E_g$ is illustrated in the insert of Fig. 2. 

\begin{figure}[hbp!] 
\centerline{\epsfig{file=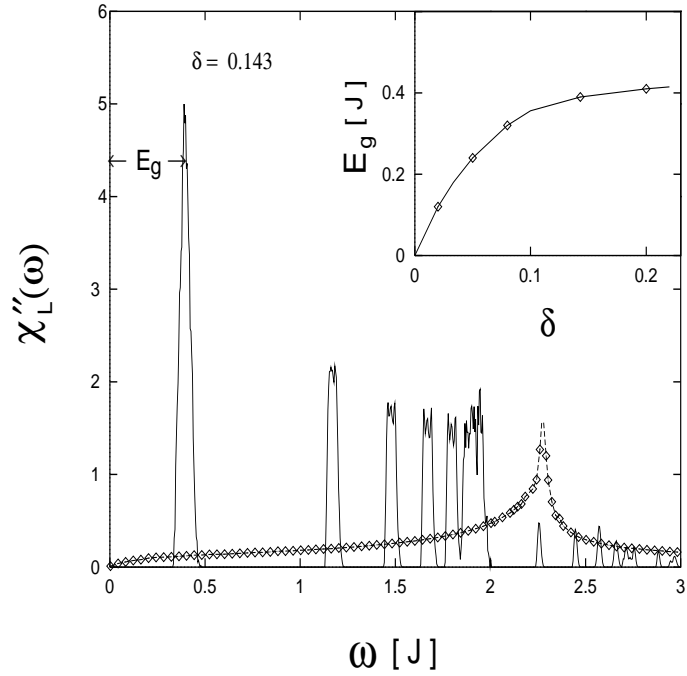,height=3.5in,width=3.5in}}
\vspace{10pt}
\caption{The local dynamic spin susceptibility function versus energy at $\delta=0$ ($\diamond$ curve) and $\delta=0.143$ (solid curve). The insert:
$E_g$ versus the doping concentration. }
\label{fig2}
\end{figure}

Therefore, in the bosonic RVB state where the AF correlations are well described, the superconducting
transition temperature is essentially decided by the low-lying spin characteristic energy. According to Fig. 2, $J\sim 100$ $meV$ gives rise to
$E_g=41$$meV$ at $\delta=0.15$ which are consistent with the neutron-scattering 
data for such a compound\cite{o7}. Then at the same $E_g$, one finds $T_c\sim
100 $$K$ according to Fig. 1 which is very close to the experimental number in 
the optimal-doped $YBCO$ compound. 

The overall picture goes as follows. The bosonic RVB order parameter $\Delta^s$ controls the short-range spin correlations 
which reflects the ``rigidity'' of the whole phase covering both undoped and doped regimes,
superconducting and normal (strange) metallic states. On the other hand, $T_c$ is basically determined 
by the phase coherence: for those preformed RVB pairs to become true superconducting condensate, the
extra phase frustration introduced by doping has to be suppressed. Here
we see how the AF fluctuations and superconductivity interplay: the former in a form of RVB pair-breaking fluctuations causes strong frustration on the charge
part through the phase string effect and its energy scale thus imposes an upper 
limit for the transition temperature of the latter. It is interesting to see
that AF fluctuations and superconducting condensation do compete with each 
other, although the driving force of superconductivity already exists in the 
Mott antiferromagnet in a form of RVB pairing.

To summarize, even though it is very natural for a RVB pairing description of 
the Mott-insulating antiferromagnetism to develop a superconducting condensation 
in the neighboring metallic regime, the issue of superconducting transition 
temperature is highly nontrivial as the AF fluctuations in a form of RVB pair-breaking process are the {\it key} effect to scramble the phase coherence of the
superconductivity. The underestimated AF fluctuations in a fermionic RVB state 
are the essential reason causing an overestimate of $T_c$ in the same system. We 
pointed out that by starting with a {\it  bosonic} RVB description where both 
the long-range and short-range AF correlations can be accurately 
described, the AF fluctuations can effectively reduce $T_c$ to a reasonable 
value through the phase string effect controlling the phase coherence of the superconducting order parameter.  

\acknowledgments

This talk is based on a series of work done in collaboration with D. N. Sheng and C. S. Ting. I would like to acknowledge the support by the 
Texas ARP program No. 3652707 and the State of Texas through the Texas Center 
for Superconductivity at University of Houston.

\end{document}